\documentstyle[prl,aps,floats,psfig]{revtex}   
\begin{document}
\twocolumn[
\hsize\textwidth\columnwidth\hsize\csname@twocolumnfalse\endcsname

\title{Spin-Voltaic Effect and its Implications} 
\author{Igor \v{Z}uti\'{c}}
\address{Condensed Matter Theory Center, Department of Physics, 
University of Maryland, College Park, Maryland 20742}
\author{Jaroslav Fabian}
\address{Institute for Theoretical Physics, Karl-Franzens 
University, Universit\"{a}tsplatz 5, 8010 Graz, Austria}

\maketitle
 
\begin{abstract}
In an inhomogeneously doped magnetic semiconductor, an interplay
between an equilibrium magnetization and injected nonequilibrium spin leads
to the spin-voltaic effect--a spin analogue of the photo-voltaic effect.
By reversing either the sign of the equilibrium magnetization or the 
direction of injected spin polarization it is possible to switch
the direction of charge current in a closed circuit or, alternatively,
to switch the sign of the induced open-circuit voltage.
Properties of the spin-voltaic effect can be used to perform 
{\it all-electrical} measurements of spin relaxation time and 
injected spin polarization, as well as to design devices with large
magnetoresistance and spin-controlled amplification.
\end{abstract}
%\pacs{72.25.Dc,72.25.Mk}
\vspace{0.2cm}
]
\newpage
Most of the existing applications using  spin degrees of freedom in 
electronic systems exploit the magnetoresistive effects in magnetic nanostructures
involving metals (paramagnetic and ferromagnetic) and insulators~\cite{maekawa02,prinz98}.
Considering semiconductors, on the other hand, offers flexibility in the doping and 
fabrication of a wide range of hetero- and nanostructures while the nonlinear current-voltage
($I-V$) characteristics are suitable for amplification and implementing logic.
Even though many materials, in their ferromagnetic state, can have a substantial degree 
of {\it equilibrium} carrier spin polarization, this alone is usually not sufficient for 
spintronic applications~\cite{dassarma01}, which typically require 
current flow and/or manipulation of the {\it nonequilibrium} spin (polarization).
Since the early work on shining circularly polarized light~\cite{lampel68,parsons69}
and driving electrical current~\cite{clark63} to generate nonequilibrium 
spin polarization in semiconductors, the challenge has remained to understand what
the implications are of such nonequilibrium spin~\cite{general}. 

We illustrate here the influence of nonequilibrium spin on $I-V$ 
characteristics in magnetic {\it p-n} junctions~\cite{zutic02,fabian02}
magnetic analogues of ordinary {\it p-n} junctions~\cite{ashcroft76}, and
focus on the implications of the spin-voltaic effect~\cite{zutic02,fabian02,zutic03}. 
Related phenomena of the spin-charge coupling~\cite{silsbee80,johnson85}
were introduced in metallic heterostructures by Silsbee and Johnson, 
following the theoretical proposals for electrical spin injection of Aronov and Pikus
\cite{aronov76a,aronov76b} (a recent account of electrical
spin injection and a detailed list of references is given in~\cite{rashba02,takahashi03}).
In the context of semiconductors there is a long tradition of optically 
generating  (for example, by shining circularly polarized light) 
spin-polarized carriers and spin-dependent electromotive force (spin emf)~\cite{dyakonov71}, 
reviewed in~\cite{orientation84,ivchenko97}.
Both incoherent~\cite{ganichev01,ganichev02} and coherent~\cite{bhat00,stevens02}
optical generation of spin currents have been demonstrated. In an applied magnetic
field and using quantum point contacts, electrical spin injection and detection in
semiconductor quantum dots~\cite{potok02} have been shown to have properties similar 
to spin-charge coupling~\cite{silsbee80,johnson85}.

\begin{figure}
\centerline{\psfig{file=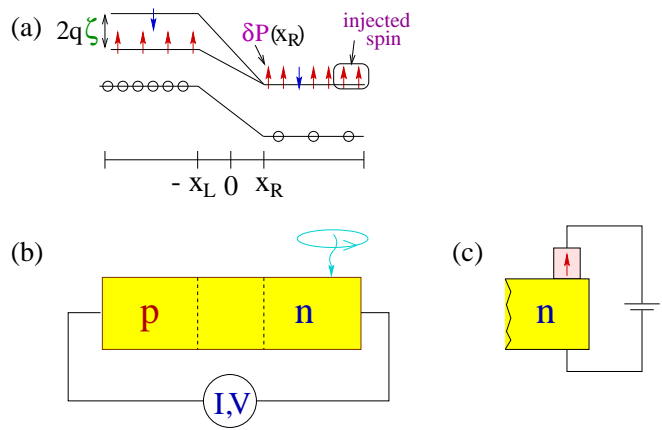,width=1.0\linewidth,angle=-90}}
\vspace{0.4truecm}
\caption{Scheme of a magnetic {\it p-n} junction. (a) Band-energy diagram with
spin-polarized electrons (arrows) and unpolarized holes (circles).
The spin splitting $2q\zeta$, the nonequilibrium spin polarization at the depletion
region edge $\delta P(x_R)$, and the region where the spin is injected are depicted.
(b) Circuit geometry corresponding to panel (a). Using circularly
polarized light 
(photo-excited electron-hole pairs absorb the angular momentum
carried by incident photons), nonequilibrium spin is injected transversely in the
nonmagnetic $n$ region and the circuit loop for  $I-V$ characteristics is indicated.
Panel (c) indicates an alternative scheme to electrically inject spin into the  
$n$ region.
}
%\vspace{-0.2truecm}
\label{fig:1}
\end{figure}

Magnetic {\it p-n} junctions, which could demonstrate the spin-voltaic effect,
have spatially dependent spin splitting of carrier bands--a consequence of 
doping with magnetic impurities and/or an applied magnetic field $B$.
Such a spin splitting can be realized using 
ferromagnetic semiconductors~\cite{munekata89}
or paramagnetic semiconductors in finite $B$ 
(where either the magnitude of $B$ or the $g$-factor is spatially 
inhomogeneous).  For simplicity, we address here the second realization,
depicted in Fig.~\ref{fig:1}.  
%fig1

In the low injection regime
it is possible to obtain the results  for
spin-polarized transport analytically and to decouple the contribution of electrons
and holes~\cite{fabian02}.
In the $p$ ($n$) region there is a      
uniform doping with $N_a$ acceptors ($N_d$ donors). Within the  depletion
region ($-x_L < x < x_R$), we assume that there is a spatially dependent
spin splitting of the carrier bands.  Zeeman splitting of the conduction band  
can be expressed as $2 q\zeta=g\mu_B B$, where $g$ is the $g$-factor for electrons,
$\mu_B$ is the Bohr magneton, $q$ is the proton charge, and
$\zeta$ is the electron  magnetic potential~\cite{zutic02}.  
In contrast to metals, the magnitude of the $g$-factor can be significantly
different from the free electron value, exceeding  500 in  
Cd$_{0.95}$Mn$_{0.05}$Se at low temperature~\cite{dietl94} and attaining  
$\approx$ 50 in InSb at room temperature. 
We further assume that the carriers obey the nondegenerate Boltzmann statistics
and consider only the effect of spin-polarized electrons
(the net spin polarization of holes can also be simply included~\cite{fabian02}).
The product of electron (n) and hole (p) densities 
in equilibrium (denoted with subscript ``0'') is  modified from the 
nonmagnetic {\it p-n} junction as~\cite{zutic02}
\begin{equation}
n_0 p_0=n_i^2\cosh(\zeta/V_T)
\label{mass}
\end{equation}
where $n_i$ is the
intrinsic (nonmagnetic) carrier density and $V_T=k_B T/q$, with $k_B$ 
the Boltzmann constant and $T$ temperature. 
In a nonmagnetic {\it p-n} junction charge current $J$ can be decomposed into
electron and hole contributions  $J=J_n+J_p$, which in turn are proportional
to the density of the nonequilibrium minority carriers, 
$J_n \propto \delta n_L=n-n_0=n_0[\exp(V/V_T)-1]|_{x=x_L}$, 
$J_p \propto \delta p_R=p-p_0=p_0[\exp(V/V_T)-1]|_{x=x_R}$,
evaluated at the two edges of a depletion region~\cite{ashcroft76},
and $V$ is the applied bias. For a magnetic {\it p-n} junction,  
Eq.~\ref{mass} implies that in the regime $\zeta > V_T$, 
the density of minority electrons changes exponentially with $B$ ($\propto \zeta$) 
and can give rise to exponentially large magnetoresistance~\cite{zutic02,fabian02}. 
Furthermore in the {\it n-p-n} magnetic bipolar transistor~\cite{fabian03}, 
from applying Eq.~\ref{mass} to the base $p$-region, it follows that the current 
amplification also varies exponentially with $B$.

Before discussing the implications of the injected nonequilibrium spin 
in a magnetic {\it p-n} junction (depicted in Fig.~\ref{fig:1}),
it is helpful to recall two simpler situations. Consider first
the usual photo-voltaic effect in a nonmagnetic {\it p-n} junction
($\zeta\equiv0$) illuminated entirely by unpolarized light.
Photo-generated electron and holes will be swept away in  opposite
directions by the built-in electric field in the depletion region.
This departure from the equilibrium carrier densities (prior to the illumination) 
shifts the balance of the electron and hole currents which no longer add up 
to zero. If the leads are connected to the two ends of a {\it p-n} junction, 
a reverse charge current will flow. Conversely for an open circuit 
configuration, photo-generated carriers suppress the built-in field 
and create a net (open circuit) voltage measured at the two terminals. An
analogous illumination by circularly polarized light can serve as a  source of spin-polarized 
current--a spin-polarized solar cell~\cite{zutic01}. However, while changing the
degree of circularly polarized light (say, by reversing the helicity of an incident light)
changes the degree of current spin polarization~\cite{zutic01},
there are no changes in $I-V$ characteristics--the nonequilibrium spin
does not affect the charge properties. 
 
\begin{figure}
\centerline{\psfig{file=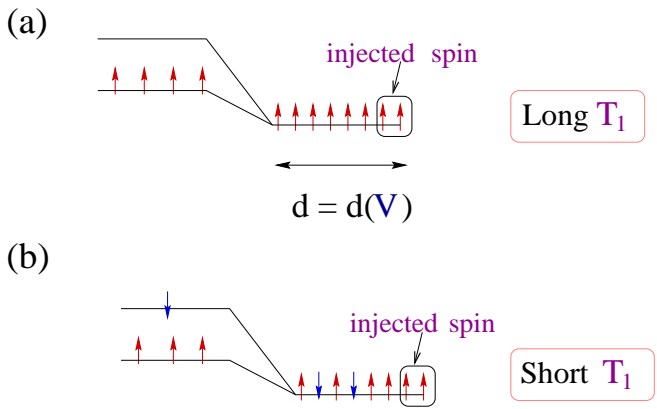,width=1.0\linewidth,angle=-90}}
\vspace{0.4truecm}
\caption{Schematic representation of a conduction band 
in a magnetic {\it p-n} junction. Spin relaxation 
of the injected spin is depicted in the limits of (a) long
and (b) short spin relaxation time $T_1$. Applied bias $V$
changes the width of the depletion region and therefore
changes the effective length between the injected spin and
the depletion region edge.}
\label{fig:2}
\vspace{-0.4cm}
\end{figure}

In a magnetic {\it p-n} junction carrier spin polarization $P=s/n$, the ratio of 
the spin $s=n_\uparrow-n_\downarrow$ and electron density ($n=n_\uparrow+n_\downarrow$),
can be changed from the  equilibrium value $P_0=\tanh(\zeta/V_T)$
by shining circularly polarized light or by direct electrical spin injection
as depicted in Fig.~\ref{fig:1} (a) and (b), respectively. With the spatially
dependent spin splitting, charge current will acquire an additional 
component--the spin-voltaic current $J_{sv}$ caused by the nonequilibrium 
spin~\cite{zutic02}. 
In the spin-voltaic effect the nonequilibrium spin, by diffusion (and drift)
from the point where it is generated
to the edge of depletion region [$\delta P_R \neq 0$, see Fig.~\ref{fig:1}(a)], 
disturbs the balance between the generation and recombination currents 
(Fig.~\ref{fig:1})~\cite{zutic02,fabian02}.
%The physics of the spin-voltaic effect is that nonequilibrium spin, diffusing 
%to the edge of depletion region [$\delta P_R \neq 0$, see Fig.~\ref{fig:1}(a)] from 
%the point where it is generated disturbs the balance between the generation and
%recombination currents (Fig.~\ref{fig:1})~\cite{zutic02,fabian02}.
If $\zeta > 0$, and more spin up electrons are
present at $x_R$ ($\delta P_R > 0$), the barrier for them to cross the region
is smaller than the barrier for the spin down electrons, so more electrons
flow from $n$ to $p$ than from $p$ to $n$, and positive charge current
results. If there are more spin down electrons at $x_R$ ($ \delta P_R < 0$), 
the current is reversed. In an open circuit geometry with a reversal of injected 
spin polarization ($\delta P_R \rightarrow -\delta P_R$)~\cite{helicity}
the sign of the induced voltage will be switched \cite{fabian02}, characteristic 
also for spin-charge coupling~\cite{silsbee80,johnson85} and the electrical detection of 
injected spin in quantum dots~\cite{potok02}. An equivalent change of the
sign for $J_{sv}$ and the open circuit voltage can be realized by reversing
the equilibrium magnetization i.e. $\zeta \rightarrow -\zeta$ and, correspondingly,
$P_{0L} \rightarrow - P_{0L}$. Such a reversal can be achieved by changing 
$B \rightarrow -B$ in the paramagnetic case, or by  temporarily applying a
finite $B$ to flip the magnetization of a ferromagnetic region (in metallic
multilayers this is feasible even at small fields $<10$ G~\cite{maekawa02}).
The total charge current can be expressed as  
spin equilibrium parts $J_n$ and $J_p$ and the spin-voltaic current $J_{sv}$
generated by the nonequilibrium spin~\cite{zutic03}
\begin{equation}
J_{sv}\propto n_{0R} P_{0L} \delta P_R  e^{V/V_T},
\label{sp}
\end{equation}
which, unlike $J_{n,p}$, can remain finite even as $V \rightarrow 0$,
when $J\rightarrow J_{sv}$~\cite{zutic02}. The product of the equilibrium magnetization and the 
nonequilibrium injected spin or, equivalently, $P_{0L} \delta P_R$ also enters 
directly the  current amplification in a magnetic
bipolar transistor~\cite{fabian03}, which can be tuned by the controlling 
the spin-voltaic effect. 
Furthermore, Eq.~\ref{sp} shows the sensitivity of $J_{sv}$ to spin relaxation 
since  $\delta P_R$ depends on the effective distance $d$
between the point of spin injection and the depletion region edge $x_R$ \cite{zutic03}. 
The decay of the corresponding nonequilibrium spin can be characterized 
by the spin relaxation time $T_1$ and the corresponding length scale, the spin diffusion 
length $L_{sn}=\sqrt{D_nT_1}$, where $D_n$ is the electron diffusivity \cite{zutic01}.
Figure~\ref{fig:2} illustrates how the sensitivity of $J_{sv}$ to spin relaxation
could be used to perform all-electrical measurements of $T_1$~\cite{zutic03}. 
Consider an idealized situation where the injected spin is completely polarized.
A long $T_1$ implies that the injected spin polarization will not be reduced at
the depletion region edge, as shown in  Fig.~\ref{fig:2}(a). 
The spin up electrons will then be easily transfered across the lower barrier in
the depletion region, leading to large $J_{sv}$. In contrast, for a short $T_1$
sketched in Fig.~\ref{fig:2}(b), some of the injected spin up electrons
will have their spin flipped. Those spin down electrons would  
go across the higher barrier (suppressed by $\propto \exp(\zeta/V_T)$ within the 
Boltzmann statistics as compared to the transfer of spin up electrons)
and $J_{sv}$ is reduced. While these two limiting cases indicate that
is possible to extract an unknown $T_1$ from $J_{sv}(T_1)$ at finite bias 
the total charge current $J$ could be dominated
by $J_n$ and $J_p$--a large $T_1$-independent background ($J_{n,p}$ do not contain 
the nonequilibrium spin). It is therefore useful 
to use the symmetry properties of the individual contributions to the 
charge current
with respect to the applied magnetic field~\cite{zutic03} 
\begin{equation}
J_{n,p}(-B)=J_{n,p}(B), \: \:  \: \:
J_{sv}(-B)=-J_{sv}(B), 
\label{parity}
\end{equation}
recalling that $\zeta \propto B$.
Consequently, $I-V$ characteristics can be used to extract the $T_1$ and 
the degree of the injected spin polarization by measuring 
$J(V,B)-J(V,-B)=2J_{sv}$~\cite{zutic03}, where
the large $T_1$-independent background has then been effectively removed. 
By adapting this procedure to the parameters of GaAs it was shown that
a variation of an unknown $T_1$ by an order of magnitude would give a
change of  approximately two orders of magnitude in $J_{sv}$~\cite{zutic03} 
and the spin relaxation time could be extracted from measured I-V curves.

To simplify the presentation, we have focused on a particular implementation
of the spin-voltaic effect. Analytical results for a low injection regime,
where $V<V_T ln (N_a N_d/n_i^2)$--the built-in potential~\cite{ashcroft76},
are also available in other cases~\cite{fabian02}. For example, when both 
$p$ and $n$ regions are magnetic, the spin-voltaic current in Eq.~\ref{sp} 
should be modified by replacing $P_{0L} \rightarrow (P_{0L}-P_{0R})/(1-P^2_{0R})$.
At higher applied bias, the corresponding problem can be solved numerically
by self-consistently combining Poisson's and the appropriate continuity
equations for spin and charge densities~\cite{zutic02}. 

A materials realization of the spin-voltaic effect, similar to the usual photo-voltaic effect,
is not limited to {\it p-n} junction, and can also include heterojunctions and structures 
with Schottky barriers. Even though most of the currently studied ferromagnetic 
semiconductors~\cite{munekata89}, such as (Ga,Mn)As and (In,Mn)As, are $p$-doped
with spin-polarized electrons rather then holes, they still have a spin splitting
in the conduction band as depicted in Fig.~\ref{fig:1}(a), and could be suitable
for the implementation of the spin-voltaic effect.

We thank S. Das Sarma and H. Munekata for useful discussions.
This paper is based on the presentation at the International Workshop on 
Nanostructured Metallic Materials
sponsored by Nanotechnology Research Network Center of Japan and Tohoku
University Materials Research Center.
This work was supported by DARPA, NSF-ECS, and the US ONR.

\newpage


\begin{references}
\bibitem{general}
In general, the nonequilibrium spin is not limited just to the carriers 
(electrons and holes in semiconductors) but includes also the spin of 
excitations and nuclei.
\end{references}
\end{document}